# Bottom-up Graphene Nanoribbon Field-Effect Transistors.

*Patrick B. Bennett[1,2], Zahra Pedramrazi[3], Ali Madani[2], Yen-Chia Chen[3,6], Dimas G. de Oteyza[3,5], Chen Chen[4], Felix R. Fischer[4,6], Michael F. Crommie[3,6], Jeffrey Bokor[2,6]\**

1: Applied Science & Technology, University Of California, Berkeley, CA 94720, USA

2: Dept. of Electrical Engineering & Computer Sciences, University Of California, Berkeley, CA 94720, USA

3: Dept. of Physics, University Of California, Berkeley, CA 94720, USA

4: Dept. of Chemistry, University Of California, Berkeley, CA 94720, USA

5: Centro de Física de Materiales CSIC/UPV-EHU-Materials Physics Center, San Sebastián, E-20018, Spain

6: Materials Sciences Division, Lawrence Berkeley National Laboratories, Berkeley, CA 94720, USA

Graphene nanoribbons (GNRs) have been extensively investigated as a promising material for use in high performance, nano-electronic, spintronic, and optoelectronic devices due to their unique physical properties[1-11]. These properties, however, are critically determined by the precise geometry of the GNR and are degraded by rough edges. Bottom-up chemical synthesis has been shown to produce GNRs *en masse* that, unlike GNRs previously studied, possess uniform width and precise edge structure[12]. Previously, the electronic structure of chemically synthesized GNRs has been studied on their Au growth substrate through Raman, photoemission and tunneling spectroscopy[12-17], but their short length and the metallic growth substrate has thus far prevented standard electronic device fabrication and transport measurements. Here we report layer transfer of chemically synthesized, atomically precise GNRs, enabling study of their physical properties regardless of substrate. Further, we fabricated nanoscale field-effect transistors based on this material and report unique transport behavior characteristic of sub-1nm GNRs.

Growth of GNRs, as previously reported[12], occurs via a two step process in which the molecular precursor, 10,10'-dibromo-9,9'-bianthryl (DBBA) is thermally sublimed in ultrahigh vacuum (UHV) onto Au(111), where it is converted into a polymer chain. Thermal cleavage of the labile C–Br bonds induces a radical step growth polymerization to yield polymeric GNR precursors. Annealing these polymers on the surface leads to a stepwise cyclization/dehydrogenation sequence yielding fully conjugated GNRs with atomically defined armchair edges. GNRs synthesized with DBBA are exactly 7 carbon atoms across (n=7, $w$=7.4 Å) with a band gap on Au(111) of approximately 2.5 eV[13-15,18].

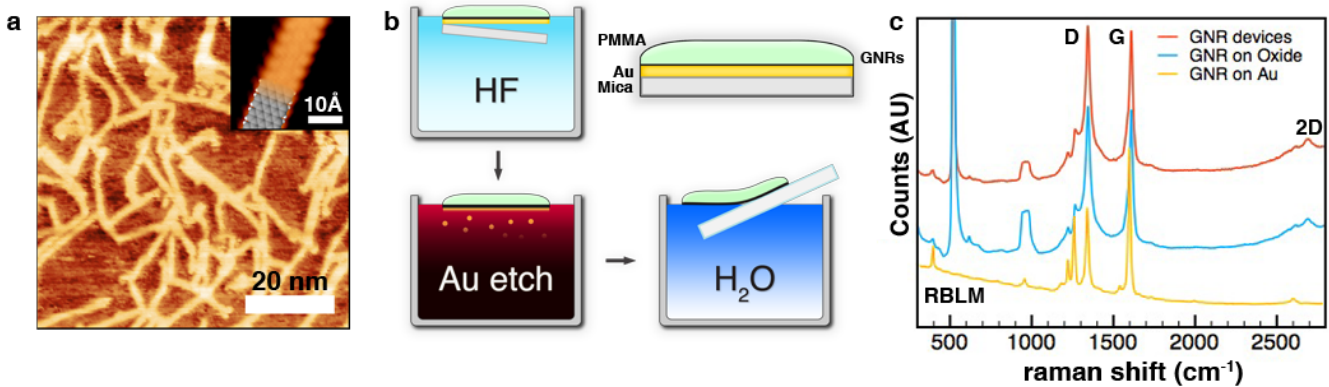

Fig. 1: **Growth and Transfer of GNRs.** (a) Room-temperature STM image of n=7 armchair GNRs on their Au growth substrate, tunneling current $I_t$ = 0.10 nA, sample bias $V_s$ = 1.67 V. Inset: high resolution image of n=7 GNR acquired with a low-temperature STM (T = 7 K, $I_t$ = 0.26 nA, $V_s$ = −0.40 V). A structural model is overlaid on the STM image. (b) Illustration of transfer process. The PMMA/GNR/Au/Mica stack is first floated on HF to delaminate the mica substrate. It is then rinsed and placed on Au etchant to dissolve the catalyst layer. It is then rinsed again and pulled onto the target substrate (c) Raman spectra of GNRs on growth substrate, after transfer on $SiO_2$, as well as after device fabrication. Peaks characteristic of n=7 GNRs are labeled for reference.

Synthesis takes place on the crystalline terraces of clean epitaxial Au films pre-deposited on cleaved mica substrates. Device fabrication requires the transfer of GNRs to an insulating substrate. The transfer process we have developed is illustrated in figs. 1(a) and 1(b) (see methods for full details). First, poly-methyl methacrylate (PMMA) is spun-cast onto the GNRs, forming a PMMA/GNR/Au/Mica stack. The stack is then floated on concentrated HF, which induces the mica substrate to delaminate from the Au growth layer. The PMMA/GNR/Au film is then rinsed and transferred to Au etchant. Finally, the PMMA/GNR film is rinsed again and drawn onto the target substrate surface, 50 nm thick $SiO_2$ thermally grown on heavily doped silicon in this study. Once the film is adhered to the substrate, it is baked to remove residual water and stripped of PMMA with acetone, leaving GNRs on an insulating surface. Raman Spectroscopy performed on samples pre- and post- transfer verifies ribbon integrity is maintained throughout the transfer and device fabrication processes (fig. 1c), confirmed by preservation of the radial breathing like mode (398 cm$^{-1}$) characteristic for n=7 GNRs. An observed increase in the D peak intensity (1343 cm$^{-1}$) and slight overall linewidth broadening may be the result of reduced substrate screening effects[19] or defects induced during transfer. This transfer process is also compatible with any substrate resistant to organic solvents such as acetone.

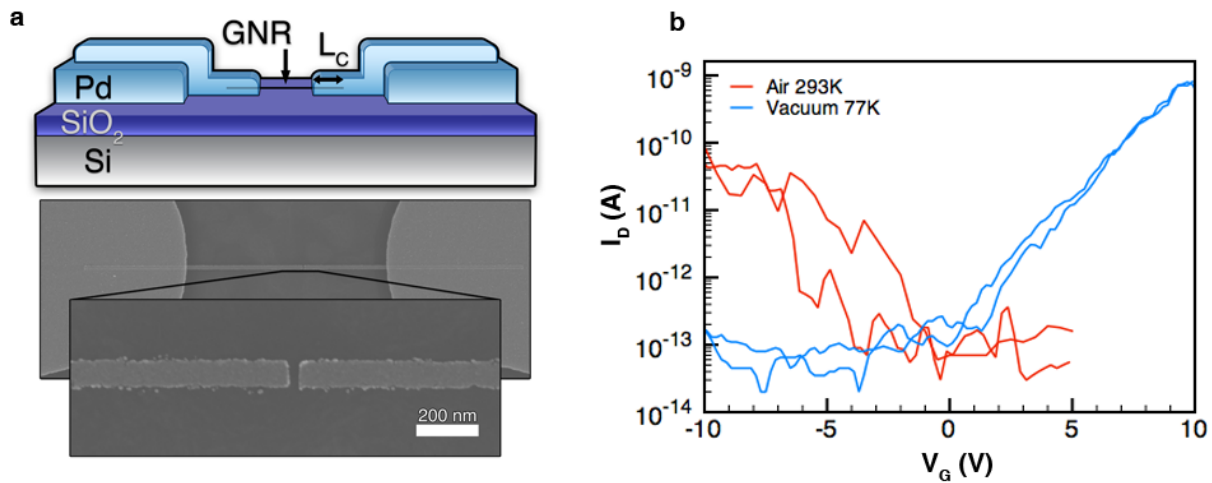

Fig. 2: **Device Fabrication and Environmental Behavior** (a) Schematic illustrating device geometry. Because small channel lengths are necessary, a Pd layer forming source and drain contacts to the GNR, using e-beam lithography, is connected to an optically defined Pd layer used to pattern contact pads. The GNR spans both contacts with some overlap region, $L_C$, between the GNR and contact. Below: Scanning electron micrograph (1keV) of the device presented in fig. 3, with 100nm wide, 26nm source drain gap. (b) Electrical characterization of a typical device at $1V_{SD}$ in both air and under vacuum at 77K.

We next fabricated three terminal transistor devices (see methods). While our GNRs can be as long as 30-40nm, the average length is 10-15nm and so very short physical channel lengths are necessary to contact ribbons at both the source and drain. Therefore, to measure individual ribbons, source and drain contacts with nanoscale gaps and 100nm width were defined using e-beam lithography (fig. 2a).

Devices fabricated with patterned source-drain gaps greater than 30nm do not show any conductance, implying that possible inter-ribbon charge transfer between any overlapping ribbons is negligible and that single GNRs did not directly bridge any source-drain gaps this wide. Several devices with smaller gaps between 20-30nm (14 out of 300 devices, gaps ranging 20-40nm) exhibit gate-modulated conductance with on-currents ranging from tens of pA to a few nA at 1V source-drain bias, $V_{SD}$. Because ribbon orientation and position is random, the actual channel length and number of ribbons in each individual device is uncertain. We estimate that in each device there are zero to two GNRs long enough to potentially contact both the source and drain; GNR density is approximately $2\times10^4/\mu m^2$ with less than 4% of ribbons longer than 30nm. Device yield is expected to increase significantly by further scaling the source-drain gap and/or increasing ribbon length during synthesis.

Fig. 2b presents electrical characterization of a typical GNR transistor measured in ambient conditions (red) and under vacuum at 77K (blue). When measured in air, GNRs contacted with Pd exhibit p-type conduction. Immediately post-fabrication, transistors exhibit large random conductance variations and variable hysteresis due to adsorbed oxygen, water, and residual PMMA on the contact and GNR[20,21]. Once annealed in vacuum, device behavior switches to n-type conduction, caused by reduction of the contact metal work function due to molecular desorption[22], and hysteresis is greatly reduced by desorption from the channel. About half of devices still display hysteretic ambipolar behavior after vacuum annealing or re-exposure to ambient conditions. Further passivation with a hydrophobic monolayer, hexamethyldisilazane (HMDS), was found to nearly eliminate hysteresis and fully switch device polarity in all devices. Residual hysteresis effects are attributed to trapped charges within the relatively thick back-gate dielectric and not from molecular adsorbates on the contact or channel[23].

Transport is largely dominated by the Schottky junction contacts. Full polarity switching through small shifts in contact work function[24], relative to the GNR's ~2.5eV band-gap, suggests that band alignment of the Pd Fermi level falls close to mid-band-gap, a conclusion in agreement with simulations of n=7 GNR/Pd interfaces[25]. Previously published experiments measuring the electrical characteristics of unzipped, chiral Pd contacted GNR transistors 2-20nm wide, derived from CNTs through sonochemical exfoliation, observe the presence of a relatively small Schottky barrier at the metal-GNR interface[5,7]. We see much larger Schottky barriers, potentially as large as 1.25 eV, due to the increased band-gap of the much narrower (7.4 Å) GNRs.

Fig. 3 presents full transport characterization of another typical device, with 26nm source-drain gap, as a function of back gate modulation (fig. 3a) and source-drain bias (fig. 3b). High series resistance is also presumed to limit on-current in our GNR devices due to the short contact overlap length ($L_C$) between the GNR and the source and drain. Even with our very short channel gaps, contact between the GNRs and Pd is no more than a few to perhaps 10 nm long depending on GNR length and alignment. Conventional graphene and CNT transistors show large resistance increases as $L_C$ is decreased past the electron mean free path ($\lambda$~200nm) [26,27], suggesting very low transmission probabilities in short contacts for our smooth edged GNRs with low scattering. Despite this, we still see a large measurement limited on-off ratio of $3.6 \times 10^3$ at $V_{SD} = 1V$, clearly demonstrating semiconducting transport in chemically synthesized GNRs.

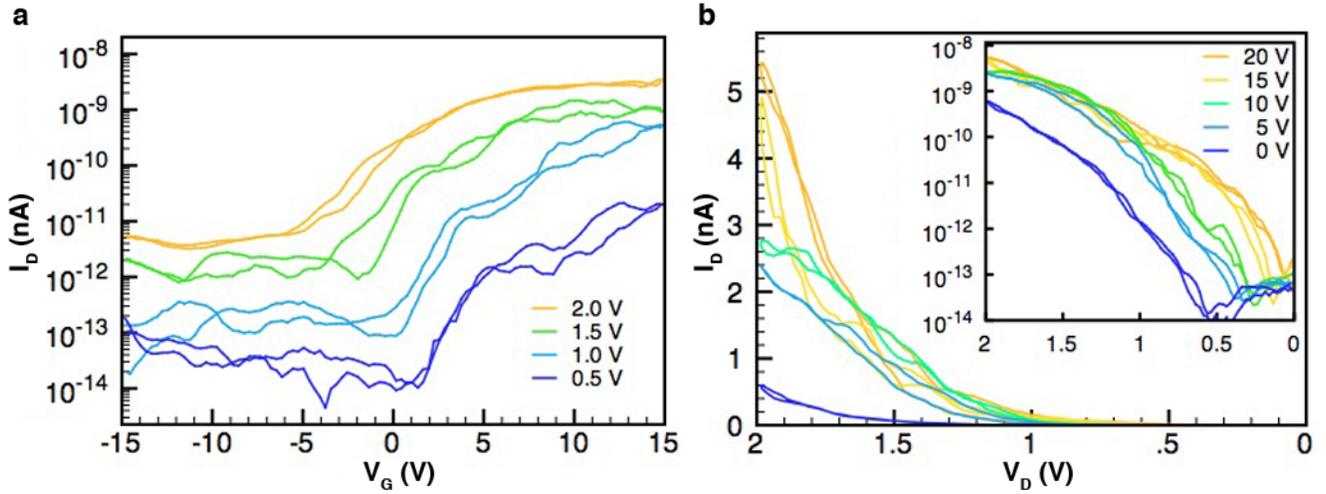

Fig. 3: **Electrical Characterization of a typical device post passivation, under vacuum, at 77K.** (a) drain current response with respect to gate voltage, $I_D$-$V_G$, at different source drain bias, $V_{SD}$, and (b) drain current response with respect to drain voltage, $I_D$-$V_D$, of same device at different gate bias, $V_G$, inset: The same data presented in logarithmic scale.

The observed device behavior is typical of a short channel Schottky barrier device[28]. In the off-state, leakage is caused by holes tunneling through the drain barrier, which is therefore relatively temperature independent but strongly dependent on $V_{SD}$, as larger biases will narrow the width of the Schottky barrier substantially. Also strongly dependent on $V_{SD}$ is the threshold voltage ($V_T$) for turn-on, becoming negative for $V_{SD}$>1V, due to the strong coupling of the channel to the drain that the gate has to counteract for the device to remain off. The large electric field between the source and drain when the devices are gated on or aggressively biased is sufficient to induce tunneling through the barriers causing field emission to become the dominant current source. This results in unsaturated, nearly exponentially increasing on-current (fig. 3b, inset), even at large $V_{SD}$, as the barrier continues to narrow and tunneling increases. From this, we can conclude that the resistance of the GNR channel is much lower than the Schottky barrier series resistance, but intrinsic GNR transport properties cannot be observed until these extrinsic factors are ameliorated.

Lowering of the contact work function would reduce the source conduction band barrier height, correspondingly increasing the drain valence band barrier, resulting in both improved on- and off-state performance. Further improvement should also arise through the use of wider, chemically synthesized GNRs such as those recently synthesized via similar methods with 1.4nm width and ~1.4eV band-gap[18]. These are expected to show improved characteristics in a given device due to smaller Schottky barriers and lower effective mass that result from their smaller band-gap. Longer GNRs, through synthesis optimization, may also reduce contact resistance by increasing $L_C$.

The narrow width, chemically synthesized GNRs studied here appear to be more sensitive to their environment compared to graphene, CNTs, or significantly wider GNRs previously studied, possibly a consequence of a higher proportion of the exposed, current carrying edge region[14,18], relative to the chemically inert surface[29]. Novel sensors with greater sensitivity than seen with graphene or CNTs might be achieved through edge modification. Similarly, artificially induced edge states in graphene have been shown to be beneficial to graphene-metal contacts[30] and may also be engineered to enhance GNR-metal electronic coupling. Electronic behavior may also be adjusted through local environment modification in addition to precursor selection during synthesis.

By developing a method for layer transfer of chemically synthesized GNRs we have gained the ability to directly study, using techniques previously unavailable, the behavior of this bottom-up engineered, self-assembled electronic material. In addition to electronic transport measurements, other experiments using chemically synthesized GNRs are also now possible, including optoelectronic and spintronic studies, optical fluorescence measurements, or transmission electron microscopy of freestanding GNRs suspended by patterned membranes. This work highlights the materials development path toward future electronic devices with low series resistance and high intrinsic mobility expected of chemically synthesized GNRs with atomically smooth edges.

**Methods**

**Growth & Transfer Processing:**

PMMA is spun cast (4krpm) onto the substrate and baked (180°C, 10 min.). The sample is floated on HF (40% wt.) and is occasionally agitated until the Mica substrate delaminates from the Au. The sample is then scooped out and rinsed in DI $H_2O$ twice. Next the sample is floated on Au etchant ($KI-I_2$) until visibly the Au film is removed. The sample is then rinsed twice again and pulled onto the substrate. The substrate then undergoes a two-step bake process (50°C for 5 min., 100°C for 10 min.) and is then stripped of the PMMA in acetone overnight.

**Device Fabrication:**

Optical Pads are patterned using optical lithography and Pd contacts (30nm thick) are deposited via e-beam evaporation. E-beam lithography is used to expose PMMA (950K $M_w$, 4krpm spin, 180°C bake for 10 min.) in a CRESTEC CABL-9510CC lithography tool. The sample is developed at -4°C in a 7:3 H2O:IPA co-solvent solution. Pd source & drain contacts (10nm thick) are again evaporated and lifted off in 80°C acetone for ½ hr.

**Characterization & sample modification:**

STM images of as-grown 7-GNR/Au(111) samples were acquired at room temperature using a commercial Omicron variable temperature STM (VT-STM) under UHV conditions. A home-built low-temperature STM (LT-STM), operating at 7 K and under UHV, was also used to obtain high-resolution images n=7 GNRs. Raman measurements were taken using a Horiba ARAMIS scanning Raman tool with a 532nm laser. Devices were screened with an Agilent B1500A parameter analyzer in a Cascade Summit probe station and measured in a Lakeshore CRX cryogenic probe station. Samples were modified throughout device measurement. Samples were annealed at 300°C for 72 hours in vacuum (3e-7 Torr), followed by a secondary anneal in the vacuum probe station pre-measurement (80°C, 24 hours, 1.5e-6 Torr). HMDS passivation was carried out under vacuum (<10Torr $N_2$) in a YES-5 Vapor priming oven.

**Acknowledgements**

Research was supported by the Office of Naval Research BRC Program. Work at the Molecular Foundry was supported by the Office of Science, Office of Basic Energy Sciences, of the U.S. Department of Energy under Contract No. DE-AC02-05CH11231. All devices were fabricated in the UC Berkeley Nanolab. We would like to thank Prof. M. Lundstrom, Prof. Sumon Datta, Dr. D. Haberer, & Prof. S.J. Choi for useful discussions.